\documentclass[prd,showkeys,floatfix,nofootinbib,
               preprint,12pt,
               tightenlines,fleqn]{revtex4} 

\usepackage{amsmath,amssymb,revsymb,graphicx,dcolumn}
\usepackage{leftidx}

\newcommand{\beq}{\begin{equation}}
\newcommand{\eeq}{\end{equation}}
\newcommand{\beqa}{\begin{eqnarray}}
\newcommand{\eeqa}{\end{eqnarray}}
\newcommand{\bsubeqs}{\begin{subequations}}
\newcommand{\esubeqs}{\end{subequations}}

\begin{document}
%
\title[]
      {Observing quantum gravity in asymptotically AdS space\vspace*{5mm}}
\author{Slava Emelyanov}
\email{viacheslav.emelyanov@physik.uni-muenchen.de}
\affiliation{Arnold Sommerfeld Center for
Theoretical Physics,\\
Ludwig Maximilian University (LMU),\\
80333 Munich, Germany\\}

\begin{abstract}
\vspace*{2.5mm}\noindent
A question is studied whether an observer can discover quantum gravity being
in the semi-classical regime. It is shown that it is indeed possible to probe a certain quantum gravity 
effect by employing an appropriately designed detector. The effect is related to the possibility
of having topologically inequivalent geometries at once in the path integral approach.
\\[1mm]
A CFT state which is expected to describe the eternal AdS black hole in the large $N$ limit is discussed.
It is argued under certain assuptions that the black hole boundary should be merely a patch of the entire 
AdS boundary. This leads then to a conclusion that that CFT state is the ordinary CFT vacuum restricted 
to that patch. If existent, the bulk CFT operators can behave as the ordinary semi-classical quantum field 
theory in the large $N$ limit in the weak sense. 
\end{abstract}



\maketitle

\section{Introduction}

In the path integral approach to (Euclidean) quantum gravity, one integrates over all metrics including 
topologically inequivalent ones. In the saddle-point approximation one at the outset  chooses 
geometries $\{\mathcal{M}_i\}$ which are solutions of the gravity field equations 
with a prescribed boundary condition at $\partial\mathcal{M}$ being common for all of $\{\mathcal{M}_i\}$. 
If one of the geometries cannot be regular without ascribing to it an intrinsic temperature, then the rest of 
the geometries have also to be ``thermalized" for all of them to fit inside the boundary $\partial\mathcal{M}$.
This in turn leads to the concept of a thermodynamically preferred geometry among of $\{\mathcal{M}_i\}$
as well as phase transitions (topology changes) if one imposes the ordinary conditions of the thermodynamic 
stability~\cite{Gibbons&Hawking2,Hawking3}.

A particular example is provided by geometries being asymptotically anti-de Sitter (AdS) space with 
a boundary $\partial\mathcal{M}$ corresponding to compactified Minkowski space. The well-known 
geometries with these properties are anti-de Sitter ($\mathcal{M}_1$) and Schwarzschild--AdS 
($\mathcal{M}_2$) spaces. The Euclidean section of Schwarzschild--AdS space is singular on its horizon,
unless the Euclidean time is in the circle $\mathbf{S}$, where its circumference $\beta$
is inverse of the Hawking-Page temperature~\cite{Hawking&Page}. The thermodynamically 
favored phase is either $\mathcal{M}_1$ with a thermal gas or $\mathcal{M}_2$ depending on a size
of the black hole horizon. The Schwarzschild--anti-de Sitter geometry is thermodynamically stable when the 
black hole is sufficiently large. At an intermediate size of the black hole $\mathcal{M}_2$ becomes unstable 
and $\mathcal{M}_1$ geometry filled with the thermal gas turns out to be thermodynamically preferred. 
The process of the AdS black hole nucleation is known as the Hawking-Page transition from $\mathcal{M}_1$ to 
$\mathcal{M}_2$~\cite{Hawking&Page}. Dynamically it is characterised by an appearance of the so-called negative
mode in the metric perturbation which leads to instability of the geometry~\cite{Prestidge}.

The thermal gas in $\mathcal{M}_1$ is supposed to be due to the thermal gravitons and scalar field
excitations. In what follows, the gravitons are not treated. The scalar field quanta are defined with respect
to the global time-like Killing vector $K_1$. The same scalar field in the geometry $\mathcal{M}_2$ causes 
the black hole evaporation through the Hawking radiation. This radiation is usually interpreted as being due 
to the pair particle production near the horizon, such that one of the particles escapes to the infinity (if massless). 
These excitations are defined with respect to the time-like Killing vector $K_2$ (a generator of the Schwarzschild time
translation). Therefore, one may consider 
the thermal scalar field quanta as characterising the phases instead of spaces with the compactified Euclidean time.

This seems natural if one imagines an observer being near the boundary $\partial\mathcal{M}$. 
As it is pointed out in~\cite{Hawking2},
the phase transition between $\mathcal{M}_1$ and $\mathcal{M}_2$ can be thought of as a
scattering process. Essentially it means that
a detector or a scalar field observable could be sensitive to the global structure of spacetime. This occurs 
due to the global nature of a quantum state. One of the purposes of this paper is to study regimes when 
both geometries are simultaneously 
relevant by the local and non-local quantum field operators. In other words, a question is posed
whether it is possible to discover quantum nature of gravity being a semi-classical observer (excluding effects
due to gravitons~\cite{Burgess}).

As well-known anti-de Sitter spacetime is not globally hyperbolic. To have a well-defined quantum theory, 
one has in particular to specify a certain boundary condition at $\partial\mathcal{M}$.
The boundary conditions imposed on the scalar field will be chosen to be either of the 
Dirichlet type or of the Neumann one. This is mostly motivated by a possibility to relate the quantum scalar 
field operator $\hat{\Phi}(x)$, where $x \in \mathcal{M}_1$ to the conformal field operator $\hat{\mathcal{O}}(y)$ 
on the boundary, where $y \in \partial\mathcal{M}$. The conformal dimension $\Delta$ of $\hat{\mathcal{O}}(y)$ 
is either $\Delta_{+} = 2$ or $\Delta_{-} = 1$ depending on the boundary condition imposed.
This duality is known in general as the anti-de Sitter/conformal field theory (CFT) 
correspondence~\cite{Maldacena1,Gubser&Klebanov&Polyakov,Witten1} and~\cite{Balasubramanian&Kraus&Lawrence,
Banks&Douglas&Horowitz&Martinec,Balasubramanian&Kraus&Lawrence&Trivedi}. In particular, 
it implies a relation between the states in CFT and the discrete spectrum of the scalar field excitations inside of 
$\mathcal{M}_1$.

The same type of boundary conditions will be imposed on the scalar field in $\mathcal{M}_2$ at $\partial\mathcal{M}$.
This is needed for a formal consistency with the boundary conditions imposed on the field in $\mathcal{M}_1$. 
The AdS/CFT correspondence is treated in $\mathcal{M}_2$ as a formal correspondence between Fock spaces 
built on the Hartle-Hawking state and a particular entangled state of two independent (commuting) CFT 
theories~\cite{Balasubramanian&Kraus&Lawrence&Trivedi,Maldacena}. 

It is worth noting that the Hawking--Page transition has been interpreted through the AdS/CFT 
conjecture~\cite{Maldacena1,Gubser&Klebanov&Polyakov,Witten1} as the 
confinement/deconfinement phase transition in the quantum chromodynamics~\cite{Witten}. 

The scalar field theory considered in this paper is described by a non-interacting, hermitian scalar field $\Phi(x)$ 
conformally coupled to gravity (this explains the values of $\Delta_{\pm}$ given above). A quantum vacuum in this 
model is completely characterized by the two-point 
function $W(x,x')$ (the $\mathbf{Z}_2$ symmetry excludes non-vanishing one-point function). 
The two-point function $W_\text{AdS}(x,x')$ in the AdS geometry, $\mathcal{M}_1$,  can be exactly computed. 
The exact expression of $W_\text{BH}(x,x')$ in $\mathcal{M}_2$ is hard to obtain if even possible. However, it 
is quite well approximated by the Gaussian two-point function $W_\text{Gau{\ss}}(x,x')$ in the case of the Einstein 
universes~\cite{Page}. To satisfy above mentioned boundary conditions, for instance, in $\mathcal{M}_{2}$, one adds 
extra term to $W_\text{Gau{\ss}}(x,x')$. As will be shown,
this leads to rather different observations of the semi-calssical observer near the boundary $\partial\mathcal{M}$.

Having the Gaussian two-point function in the background of the AdS black hole, it is straightforward then to compute it 
on the AdS boundary as well. The Hartle-Hawking state should then be equivalent to a certain CFT state, 
$|\Omega\rangle$. This is a pure high-energy CFT state which looks thermal in the large $N$ 
limit~\cite{Banks&Douglas&Horowitz&Martinec}. It is a non-trivial problem to find $|\Omega\rangle$ and
CFT operators which should correspond the bulk scalar quantum theory in $\mathcal{M}_2$.

The outline of this paper is as follows. In Section \ref{sec:Asymptotically AdS spaces}, Euclidean quantum gravity 
is briefly reviewed following~\cite{Gibbons&Hawking2,Hawking3} for the sake of completeness of the presentation. 
The correlation 
function for both types of the boundary conditions are computed in anti-de Sitter and Schwarzschild--anti-de Sitter 
spaces. In the case of $\mathcal{M}_1$, the Gaussian approximation is exact. It appropriately describes the state in 
$\mathcal{M}_2$ only near the boundary $\partial\mathcal{M}$ if one imposes either the Dirichlet or the Neumann 
boundary condition. In Section \ref{sec:Detector responses}, different types of detectors or scalar field observables 
are introduced. As will be shown, a detector that is sensitive to the integral effect of the quantum field fluctuations
could be used to reveal quantum nature of gravity. A particular example of such kind of the detector analyzed in
this paper is the Unruh-DeWitt detector. A detector of a local type is also treated. It turns out to be oblivious to
the spacetime global structure if located near the AdS boundary, but sensitive to the boundary conditions.
In Section \ref{sec:Discussion and concluding remarks}, I will discuss the main results of this paper and speculate
about nature of the CFT state which is expected to reduce to the Hartle-Hawking vacuum in the semi-classical 
approximation (or at the leading order in $1/N$).

Throughout this paper the fundamental constants are set to unity, $c = G = k_\text{B} = \hbar = 1$.

\section{Asymptotically Anti-De Sitter spaces}
\label{sec:Asymptotically AdS spaces}

In this paper, I will deal only with two asymptotically AdS spaces. These are the AdS and Schwarzschild--AdS
manifolds. The Schwarzschild--AdS line element is given by
\beqa\label{eq:metric:schwarzschild-ads-space}
ds^2 &=& g_{\mu\nu}dx^{\mu}dx^{\nu} \;=\;
f(r)dt^2 - \frac{dr^2}{f(r)} - r^2d\Omega^2\,, \quad f(r) \;=\; 1 - \frac{2M}{r} + H^2r^2\,,
\eeqa
where $M$ is the black hole mass and $H$ is the AdS Hubble parameter set to unity, $H \equiv 1$.
If one sets $M = 0$, then \eqref{eq:metric:schwarzschild-ads-space} takes the form of the AdS line element. In this
case, the lapse function $f(r)$ is denoted as $f_0(r)$.

\subsection{\bf Euclidean quantum gravity}

The Hawking-Page phase transition can be discovered when one goes beyond the semi-classical approximation, 
wherein the metric is a classical field. In the path integral approach to quantization of gravity, a primary object is
\beqa
Z &=& \int\mathcal{D}g\,\mathcal{D}\Phi\;e^{iS[g,\Phi]}\,,
\eeqa
where the integration is performed over all metrics $g(x)$ (up to diffeomorphism) including topologically 
inequivalent and field $\Phi(x)$ configurations satisfying certain boundary conditions. The path integral becomes 
more well-defined if one works with Euclidean section of spacetime, i.e. the time coordinate is purely imaginary. 
This is assumed below in this subsection. In the Euclidean space, $Z$ is then interpreted as the partition 
function~\cite{Gibbons&Hawking2,Hawking3}.

Adding a source term $J(x)$ to the scalar field Lagrangian, $J(x)\Phi(x)$, one replaces $Z$ by the
functional $Z[J]$. This functional is a convenient instrument to generate correlation functions of the scalar field. 
The vacuum being
under consideration is Gaussian, i.e. it is completely determined by a two-point correlation function (assuming
scalar field action is symmetric under $\Phi(x) \rightarrow -\Phi(x)$). If one defines
\beqa\label{eq:feynman-function}
\langle\hat{\Phi}(x)\hat{\Phi}(x')\rangle &\approx&
\sum_{i} \int\mathcal{D}\Phi\; \Phi(x)\Phi(x')\;e^{-S_\text{E}[g_i,\Phi]}\,,
\eeqa
then one chooses the Euclidean state at the outset. It usually corresponds to the so-called Hadamard state, 
i.e. it is well-defined over entire 
spacetime. In the above formula the saddle-point approximation has been used, i.e. only a countable set of manifolds 
$\mathcal{M}_i$ (with metrics $g_i(x)$, where $x \in \mathcal{M}_i$) have been taken into account at the tree level. 
Quantum nature of gravity appears here as simultaneously coexisting spaces with different topologies.

Among of all possible $\mathcal{M}_i$, one selects those which have the same boundary. The boundary of asymptotically
AdS$_4$ space has a topology $\mathbf{S}{\times}\mathbf{S}^2$. The manifolds fitting inside this boundary
are Schwarzschild-AdS space ($\mathbf{S}^2{\times}\mathbf{D}^2$) and pure AdS ($\mathbf{S}{\times}\mathbf{D}^3$),
where the boundary points are included in those spaces. The pure AdS geometry is supposed to be filled with
the thermal gas at the inverse temperature $\beta_0$ being equal to the circumference of $\mathbf{S}$.

Employing the relation between the time-ordered two point function and the Wightman one, it is straightforward
to obtain from \eqref{eq:feynman-function} the total Wightman function
\beqa\label{eq:wightman-total}
W(x,x') &\approx& \gamma_\text{BH}\;W_\text{BH}^{\beta}(x,x') 
+ \gamma_\text{AdS}\;W_\text{AdS}^{\beta_0}(x,x')\,,
\eeqa
where $\gamma_\text{BH} + \gamma_\text{AdS} = 1$ and $\gamma_\text{AdS}/\gamma_\text{BH} = \exp(-\Delta S)$, 
where $\Delta S = S_\text{AdS} - S_\text{BH}$ is the difference of the Euclidean actions. Both $S_\text{AdS}$ and
$S_\text{BH}$ are infinite. Their difference is infinite as well. To make sense of $\Delta S$, one has first to regularize 
it by cutting integration over $r$ at $r_\text{c}$. Second, one assumes a local thermal equilibrium between the AdS 
black hole and the thermal gas in AdS space at $r = r_\text{c}$. In other words, the local temperature in AdS black 
hole geometry is equal to the local temperature of the thermal gas in pure AdS space at $r = r_\text{c}$~\cite{Witten}: 
$f_0^{1/2}(r_\text{c})\beta_0 =  f^{1/2}(r_\text{c})\beta$. Thus, one pretends that there is an effective isothermal cavity of size 
$r_\text{c}$ fixed at that local temperature (see~\cite{Brown&Creighton&Mann}). One then obtains a finite result for 
$\Delta S_\text{ren}$  in the limit $r_\text{c} \rightarrow \infty$ (infinitely large cavity), i.e. 
\beqa
\Delta S_\text{ren} &=& -\pi r_+^2\frac{1 - r_+^2}{1+3r_+^2}
\eeqa
at the Hawking-Page temperature $\beta = 1/T_\text{HP}$, where $r_+$ is a size of the black hole horizon
(see below)~\cite{Hawking&Page,Witten}.

The correlation functions appearing in \eqref{eq:wightman-total} are those for Schwarzschild--AdS space and
anti-de Sitter space at inverse temperatures $\beta$ and $\beta_0$, respectively.
Thus, the total two-point function is given by a weighted sum of two correlators corresponding to the pure AdS 
and AdS black hole geometries at those inverse temperatures. The weights $\gamma_\text{BH}$ and $
\gamma_\text{AdS}$ are interpreted as probabilities of the occurrence of these AdS black hole and AdS with 
the thermal gas, respectively, in the vacuum.

\subsection{Wightman two-point functions}

It is convenient to work with the conformally rescaled metric $\bar{g}_{\mu\nu}(x) = g_{\mu\nu}(x)/f(r)$ which is
known as the optical metric. The barred quantities introduced below are defined with respect to the optical metric.

\subparagraph{\bf Anti-de Sitter geometry.}

In the limit $M \rightarrow 0$, the lapse function $f(r)$ defined in \eqref{eq:metric:schwarzschild-ads-space} 
reduces to $f_0(r) \equiv 1 + r^2$. This corresponds to AdS space. The AdS space is a maximally symmetric manifold
with the Killing algebra $\mathfrak{so}(2,3)$. The Killing vector $K_1 = \partial_t$ is identified with $L_{05} \in \mathfrak{so}(2,3)$
and used to define positive- and negative-frequency field modes associated with the AdS vacuum~\cite{Avis&Isham&Storey,
Breitenlohner&Freedman}. The Wightman functions for the Dirichlet and Neumann
conditions\footnote{Note that these conditions are imposed on the rescaled field, i.e. $\bar{\Phi}(x) = \Phi(x)f_0^{\frac{1}{2}}(r)$.
The physical field $\Phi(x)$ vanishes on the AdS boundary $\partial M$ for both of them.}
at the AdS boundary~\cite{Breitenlohner&Freedman} read

\beqa\label{eq:wightman-ads}
\overline{W}_\text{AdS}^{\pm}(x,x') &=& 
\frac{1}{8\pi^2}
\frac{1}{\cos\left(\Delta t\right) - \cos\left(r_* - r_*'\right)}
\mp \frac{1}{8\pi^2}\frac{1}{\cos\left(\Delta t\right) - \cos\left(r_* + r_*'\right)},
\eeqa
where $\Delta t \equiv t - t'$ and $\text{Im}(\Delta t) \rightarrow -0$. 
The coordinate $r_* = \arctan r-\frac{\pi}{2}$ is the Regge-Wheeler coordinate in AdS space, and the
angular coordinates of $x$ and $x'$ have been equated as being irrelevant in the following. Note that the upper 
(lower) sign in \eqref{eq:wightman-ads} and below refers to the Dirichlet (Neumann) boundary condition.

The two-point function \eqref{eq:wightman-ads} modifies if one considers the scalar field at non-zero temperature.
A quantum state describing the thermal gas of the scalar field excitations in AdS being at the inverse temperature $\beta_0$ 
corresponds to the two-point function $\overline{W}_\text{AdS}^{\beta_0}(x,x')$. This correlation function is invariant under 
$\Delta t \rightarrow \Delta t + i\beta_0$. In other words, $\overline{W}_\text{AdS}^{\beta_0}(x,x')$ is periodic in imaginary
time coordinate with a period equal to the inverse temperature $\beta_0$. This is achieved by replacing 
$\Delta t$ in \eqref{eq:wightman-ads} by $\Delta t + in\beta_0$, where $n \in \mathbf{Z}$, and then summing over all integers $n$.
The state described by $\overline{W}_\text{AdS}^{\beta_0}(x,x')$ is known as the Kubo-Martin-Schwinger (KMS) state 
defined with respect to the Killing vector $K_1 = \partial_t$~\cite{Haag}.

\subparagraph{\bf AdS black hole geometry.}

If $M$ in \eqref{eq:metric:schwarzschild-ads-space}  is non-vanishing, the field $\partial_t$ is still a Killing vector. 
It is denoted as $K_2$ in the Introduction. The vector $K_2$ is time-like only outside of the black hole. 
At the black hole horizon $r = r_+$, where $r_+$ is the real root of $f(r_+) = 0$, the vector $\partial_t$ becomes 
null and space-like for $0 \leq r < r_+$. The Boulware state is vacuous for an observer moving along $\partial_t$ 
outside the black hole as the field excitations are determined by the positive- and negative-frequency modes 
of $K_2$. The Wightman two point function that corresponds to the Boulware state at the inverse temperature 
$\beta$ in the Gaussian approximation~\cite{Page} reads
\beqa\label{eq:wightman-boulware-thermalized}
\overline{W}_\text{Gau{\ss}}^{\beta}(x,x') &=& \frac{1}{8\pi^2\rho_{-}}
\frac{\Delta^{\frac{1}{2}}(x,x')\,\kappa \sinh\left(\kappa\rho_{-}\right)}
{\cosh\left(\kappa\rho_{-}\right)-\cosh\left(\kappa\Delta t\right)}\,,
\eeqa
where $\rho_{-} \equiv (2\sigma_3)^{\frac{1}{2}}$, $\sigma_3(x,x')$ is the three-dimensional geodetic 
interval~\cite{DeWitt} for the optical metric $\bar{g}_{ij}(x)$, where the indices $i,j$ run from 1 to 3. 
$\Delta(x,x')$ is the van Vleck-Morette determinant for the optical metric~\cite{DeWitt}. The parameter 
$\kappa \equiv 2\pi/\beta$ reduces to the 
surface gravity if the field is in the Hartle-Hawking state. One can directly show at least up to the fifth 
order in the difference $(r-r')$ that $\rho_{-} = r_* - r_*'$, provided the angular coordinates of $x$ and $x'$ are 
equal (see App.~\ref{app:Geodetic interval and the van Vleck-Morette determinant}). 
The radial coordinate $r_*$ is the Regge-Wheeler coordinate in AdS black hole space, i.e. 
$dr_* = dr/f(r)$, and $r_*$ is set to vanish at the boundary ($r \rightarrow +\infty$).

The Hartle-Hawking state is the Hadamard state which is regular over the whole Schwarzschild--anti-de 
Sitter manifold including its Kruskal extension. When restricted to the outside of the black hole, $r > r_+$, 
it is a KMS state with respect to $K_2 = \partial_t$ at the Hawking-Page (HP) temperature 
$T_\text{HP} = (1+ 3r_+^2)/4\pi r_+$~\cite{Hawking&Page}. In other words, when one probes the Hartle-Hawking 
state by quantum field operators having zero support in the causally unavailable part of the black hole, the state 
appears as being thermal for both ingoing and outgoing modes, although it is perfectly pure.

The two-point function in the Gaussian approximation must reduce to the AdS correlation function when the black hole 
mass vanishes. The excitation spectrums  are rather different for $M = 0$ and $M \neq 0$. At the level
of the two-point functions this transformation is achieved by setting $\kappa = i$ in 
\eqref{eq:wightman-boulware-thermalized}.\footnote{Note that the imaginary value of the surface gravity implies
that the size of the horizon is imaginary, i.e. $r_+ = i$. This in turn implies that $\Delta S_\text{ren}$ equals $-\pi$, rather than vanishes.}
Taking into account a relation between the surface gravity and the black hole mass, one then has $M = 0$. 
The correlation function obtained from \eqref{eq:wightman-boulware-thermalized} 
in this manner corresponds to the transparent condition imposed on the field $\Phi(x)$ at the AdS boundary. Applying the 
inverse transformation to \eqref{eq:wightman-ads}, one can derive
\beqa\label{eq:wightman-boulware-thermalized-dirichlet}
\overline{W}_\text{Gau{\ss}}^{\pm,\beta}(x,x') &=&
\frac{1}{8\pi^2\rho_{-}}
\frac{\Delta^{\frac{1}{2}}(\rho_{-})\,\kappa \sinh\left(\kappa\rho_{-}\right)}
{\cosh\left(\kappa\rho_{-}\right)-\cosh\left(\kappa\Delta t\right)}
\mp \frac{1}{8\pi^2\rho_{+}} 
\frac{\Delta^{\frac{1}{2}}(\rho_{+})\,\kappa \sinh\left(\kappa\rho_{+}\right)}
{\cosh\left(\kappa\rho_{+}\right) - \cosh\left(\kappa\Delta t\right)}
\eeqa
which is
associated with the scalar field satisfying the Dirichlet (Neumann) boundary condition at the conformal AdS infinity, 
where $\rho_{+} \equiv r_* + r_*'$. 
The square root of the van Vleck-Morette determinant can be computed for sufficiently small $\rho$. For spatial points 
with equal angular coordinates, one obtains
\beqa
\Delta^{\frac{1}{2}}(\rho) &=& 1 + \frac{1}{24}\Big(2ff'' - \big(f'\big)^2\Big)\,\rho^2 + \frac{1}{24}f^2f^{(3)}\,\rho^3
\\[1mm]\nonumber
&&+\frac{1}{80}
\left(\frac{7}{72}\big(f'\big)^4 + \frac{7}{18}ff''\Big(ff''-\big(f'\big)^2\Big) +2f^2f'f^{(3)} + f^3f^{(4)}\right)\rho^4 +
\text{O}\big(\rho^5\big)\,,
\eeqa
where prime denotes a differentiation with respect to $r$ and $\rho$ is either $\rho_{+}$ or $\rho_{-}$. Note that
$\Delta^{\frac{1}{2}}(x,x')$ exactly equals $\rho/\sin\rho$ in anti-de Sitter space (if $\theta = \theta'$ and $\phi = \phi'$).

The formula \eqref{eq:wightman-boulware-thermalized-dirichlet} can also be obtained by employing the method 
of images~\cite{Kennedy&Critchley&Dowker}. To this end the space outside of the black hole, i.e. $\mathcal{M}_2$, 
is reflected with respect to the AdS boundary $\partial\mathcal{M}$. In a sense there are two black holes, $\mathcal{M}_2$ and 
$\tilde{\mathcal{M}}_2$, which share the same AdS boundary.
An image $\tilde{r}_* \in \tilde{\mathcal{M}}_2$ of a given point $r_* \in  \mathcal{M}_2$ corresponds $-r_*$
in the vicinity of the boundary. 
Thus, the correlation function satisfying the Dirichlet (Neumann) boundary condition consists of two terms, namely 
$\overline{W}_\text{Gauss}^{\beta}(\Delta t, r_* - r_*') \mp \overline{W}_\text{Gauss}^{\beta}(\Delta t, r_* - \tilde{r}_*')$.
This method leads thus to the above formula \eqref{eq:wightman-boulware-thermalized-dirichlet}. 

The expansion of $\Delta^{\frac{1}{2}}(x',x)$ should be performed up to $(x'-x)^5$. Indeed, the Gaussian
approximation gives the two-point function which is valid up to $(x'-x)^5$. In general, 
$\sigma_3(x',x) = \frac{1}{2}\bar{g}_{ij}\sigma_3^i\sigma_3^j$ and $\Delta^{-1}(\Delta\sigma_3^i)_{;i} = 3$, 
where $\sigma_3^i = \nabla^i\sigma_3$ by definition and the semi-colon denotes the covariant derivative defined 
with respect to the optical (barred) metric~\cite{DeWitt}. Therefore, the Wightman function is a solution 
of the scalar field equation if and only if $\Delta^{\frac{1}{2}}(x',x)$
satisfies it. This is the case up to the order $(x'-x)^3$ for any $f(r)$ of the form $1 - 2M/r-\Lambda r^2/3$.\footnote{This 
was shown for the Schwarzschild--Minkowski geometry in~\cite{Eftekharzadeh&Bates&Roura&Anderson&Hu}.}
The van Vleck-Morette determinant up to that order is given in 
App.~\ref{app:Geodetic interval and the van Vleck-Morette determinant}.

The correlator $W_\text{BH}^{\pm,\beta}(x,x')$ is an exact (unknown) expression for the Boulware state at the temperature $1/\beta$.
In what follows, its approximate version $W_\text{Gau{\ss}}^{\pm,\beta}(x,x')$ is used to model readings of detectors 
which are located near the boundary $\partial\mathcal{M}$.

\section{Detector responses}
\label{sec:Detector responses}

To probe a quantum state, one should design a detector constructed from local field operators
to non-trivially act on this state. By examining detector's readings (comparing them with the theoretical expectations), 
one can reveal properties of this state. Two types of detectors will be considered. One of them is non-local in time. 
It means it is sensitive to a cumulative (integral) effect in time caused by the field fluctuations at a given spatial point. 
Another is local in that sense it probes the vacuum state at a given spacetime point.

These operators belong in general to the the algebra of field observables.
In the algebraic approach to quantum field theory a primary object is the unital $*$-algebra $\mathcal{A}$ 
of field operators ascribed to a given globally hyperbolic spacetime $\mathcal{M}$ endowed with a classical
metric $g(x)$. One thus writes $\mathcal{A}(\mathcal{M})$.
The star denotes an involution (the hermitian conjugation below) which maps $\mathcal{A}(\mathcal{M})$ into itself. 
The quantum scalar field $\hat{\Phi}(x)$ is treated as a map from a set of test (compactly supported smooth) functions 
$\{f(x)\}$ to $\mathcal{A}(\mathcal{M})$. In other words, $\mathcal{A}(\mathcal{M})$ is generated by the identity 
operator $\hat{\mathbf{1}}$, $\hat{\Phi}(f)$ and non-linear combinations of
the field with its derivatives (for instance, stress tensor operator). The operator $\hat{\Phi}(f)$ is given by an integral
of $\hat{\Phi}(x)f(x)$ over the manifold, i.e. $\hat{\Phi}(f)$ is the field operator $\hat{\Phi}(x)$ averaged or smeared 
over the support of the test function. Quantum states are regarded as linear, positive and normalised functionals over 
the algebra $\mathcal{A}(\mathcal{M})$.\footnote{
Note that I do not make any difference in notations between the abstract algebra $\mathcal{A}(\mathcal{M})$ and its 
representation on a certain Hilbert space.} The observables are elements of $\mathcal{A}(\mathcal{M})$ invariant
under the involution. For more details, see, for instance,~\cite{Haag,Hollands&Wald,Khavkine&Moretti}.

The simplest element of $\mathcal{A}(\mathcal{M})$ that can be used as a detector is 
\beqa\label{eq:curly-p}
\hat{C}(f) &=& \hat{\Phi}^{\dagger}(f)\hat{\Phi}(f)\,,
\eeqa
where the dagger is the hermitian conjugation. Taking a certain test function $f(x)$, one can reduce 
$\hat{C}(f)$ to the known probes (measurements) of the quantum field fluctuations.

Indeed, one can associate \eqref{eq:curly-p} with the power spectrum of the quantum fluctuations in the spatially 
flat universe. If one sets $f(x) = a^{-3}(\eta_x)\delta(\eta_x - \eta)W_{L}(\mathbf{x})$, where
$\eta_x$ is the conformal time, $a(\eta_x)$ scale factor and $W_{L}(\mathbf{x})$ a window 
function, then the square root of $\langle\hat{C}(\eta,L)\rangle$ is of the order of the typical amplitude of the 
quantum fluctuations over $L$ at the moment of conformal time $\eta$~\cite{Mukhanov&Winitzki}. 
Its characteristic, i.e. the spectral index, is a measurable quantity. In fact, it was measured in the spectrum of the temperature 
fluctuations of the cosmic microwave background generated by the primordial quantum fluctuations during the inflationary 
stage of the evolution of the universe.

One can also set $f(x') = \exp\big(iEt' - t'^2/T^2\big)\delta(\mathbf{x}' -\mathbf{x})$. Then the vacuum expectation value
of \eqref{eq:curly-p} over $TE$ in Minkowski space can be interpreted as giving 
the expected number of particles (defined with respect to $\partial_t$) in the energy range $E \pm 1/T$ crossing 
the detector in the time $2T$~\cite{Fredenhagen&Haag}. In the limit $T \rightarrow \infty$, the value of 
$\hat{C}(f)/T$ in a given quantum state can be related to the transition rate $\dot{\mathcal{F}}(E)$ of 
the Unruh-DeWitt detector~\cite{Birrell&Davies}, i.e. 
\beqa
\dot{\mathcal{F}}(E) &\propto& \lim\limits_{T \rightarrow \infty}\Big(\langle\hat{C}(E,T)\rangle/T\Big)\,.
\eeqa
In general, the Unruh-DeWitt detector is determined by the two-point Wightman function $W(x,x')$ and can be 
interpreted as giving particle absorption per unit proper time of the detector. The transition rate 
$\dot{\mathcal{F}}(E)$ reads
\beqa\label{eq:unruh-dewitt-detector}
\dot{\mathcal{F}}(E) &=& \int_\mathbf{R}d\tau\,W\big(\tau,\mathbf{x}\big)\,e^{-iE\tau}\,,
\eeqa
where $\tau$ is the proper time associated with the detector, $E$ its intrinsic energy levels defined 
with respect to the proper time~\cite{Birrell&Davies}, and $\mathbf{x}$ its spatial position. It is worth mentioning 
that the interpretation of the transition rate in terms of particles is not always physically 
possible. An alternative interpretation as the frequency (defined with respect to the proper time) spectrum of 
the quantum fluctuations at a given spatial point seems to be more appropriate~\cite{Sciama&Candelas&Deutsch}.

A slightly different probe of the quantum fluctuations has been proposed in~\cite{Emelyanov2}. 
The advantage of that is that one can also obtain the 
Planck spectrum if a detector moves along a conformal Killing vector. For instance, this could be used in the
set-up of~\cite{Emelyanov3} to show that the quantum fluctuations appear as a thermal bath for a detector moving 
along the dilatation vector $D$ in AdS space. This is in complete agreement with the results of~\cite{Emelyanov3}.

The well-known local operator is the stress tensor of the scalar field.
Its renormalised value in the vacuum provides with the back-reaction of the quantum field on space. 
However, to avoid unnecessary complications, I will instead consider the renormalised value of the
scalar field squared in the vacuum. The quantum field squared is defined as follows
\beqa\label{eq:phi-squared}
\hat{\Phi}^{2}(x) &\equiv& \lim\limits_{x' \rightarrow x} \big(\hat{\Phi}(x)\hat{\Phi}(x') - H(x,x')\hat{\mathbf{1}}\big)\,,
\eeqa
where $H(x,x')$ is the Hadamard parametrix~\cite{Hollands&Wald}. It is a covariant and local object in the sense 
it depends only on a spacetime metric, i.e. it is state-independent. For a given quantum state determined by the 
two-point function $W(x,x')$, 
\eqref{eq:phi-squared} reduces to
\beqa\label{eq:local-temperature-2squared-detector}
\Phi^{2}(x) &=& \langle \hat{\Phi}^2(x) \rangle \;=\; \lim\limits_{x' \rightarrow x} \big(W(x,x') - H(x,x')\big)\,.
\eeqa
The mean field squared $\Phi^{2}(x)$ is usually well-defined for the Hadamard state over space even with an event 
horizon. The probe by $\Phi^{2}(x)$ gives a certain measure of the local vacuum activity of the quantum field at a given 
spacetime 
point.\footnote{It has been recently argued that $\hat{\Phi}^{2}(x)$ is the local temperature operator, i.e. the mean 
field squared $\Phi^{2}(x)$ describes readings of a real (macroscopic) thermometer in a certain quantum state 
(see~\cite{Emelyanov} and references therein). However, as shown in~\cite{Emelyanov}, this interpretation can 
loose its physical meaning in particular situations. This also can be seen in Eq.~\eqref{eq:mfs-on-boundary}, where
the mean field squared is negative near $\partial\mathcal{M}$ for the Dirichlet boundary condition.}

\subsection{\bf Transition rate $\dot{\mathcal{F}}(E)$}

At large time intervals $\Delta t$, the AdS black hole part of the total two-point function is exponentially 
small, i.e. $W_\text{BH}^{\beta}(\Delta t) \sim e^{-\beta\Delta t}$ at $\Delta t \gg |\Delta S_\text{ren}|/\beta$. 
The main contribution to \eqref{eq:wightman-total} is then due to the pure AdS term at those $\Delta t$. 
However, $W_\text{BH}^{\beta}(\Delta t)$ significantly contributes to the transition rate. 
Indeed, substituting \eqref{eq:wightman-total} in \eqref{eq:unruh-dewitt-detector}, one obtains
\beqa\label{eq:transition-rate-total}
\dot{\mathcal{F}}(E) &=& 
\frac{\gamma_\text{BH}}{2\pi}\frac{E}{\exp\left(E/T_2\right) - 1}
\left(1 \mp \Delta^{\frac{1}{2}}(2r_*)\frac{\sin\left(\Omega E/T_2\right)}{\Omega E/T_2}\right)
\\[1mm]\nonumber&&
+ \frac{\gamma_\text{AdS}}{2\pi}\sum\limits_{n = 1}^{+\infty}\frac{T_1}{\exp\left(n\right) - 1}
\left(n \mp \frac{\sin\left(2nr_*\right)}{\sin\left(2r_*\right)}\right)\delta\big(n - E/T_1\big)\,,
\eeqa
where $T_1 = 1/(\beta_0 f_0^{1/2}(r))$ and $T_{2} = 1/(\beta f^{1/2}(r))$ are the local temperatures in
$\mathcal{M}_1$ and $\mathcal{M}_2$, respectively, and $\Omega = 2r_*/\beta$ by definition.

The black hole part of the total transition rate \eqref{eq:transition-rate-total} deserves a special treatment. 
If one does not impose the Dirichlet or Neumann boundary conditions, then the second term inside the brackets is 
absent. This would correspond to the transition rate like in Schwarzschild space, but at the local Hawking 
temperature~\cite{Sciama&Candelas&Deutsch}.
The Dirichlet boundary condition corresponds to the upper sign in \eqref{eq:transition-rate-total}. In this case,
$\dot{\mathcal{F}}_\text{BH}(E)$ is less than zero for sufficiently small energy $E$, i.e. there exists a critical value of the energy 
$E_\text{c} > 0$, such that $\dot{\mathcal{F}}_\text{BH}(E) \leq 0$ for $E \leq E_\text{c}$. This occurs, because 
\beqa\label{eq:square-root-delta}
\Delta^{\frac{1}{2}}(\rho) &\approx& 1 - \Big(\frac{\Lambda}{18} - \frac{M\Lambda}{3r} + \frac{M}{3r^3}
-\frac{M^2}{2r^4}\Big)\rho^2 \;\geq\; 1 
\eeqa
in asymptotically anti-de Sitter space, wherein the cosmological constant $\Lambda$ is less than zero. The Neumann 
boundary condition corresponds to the lower sign in \eqref{eq:transition-rate-total}. In this case, the transition rate
$\dot{\mathcal{F}}_\text{BH}(E)$ is positive for any value of the energy $E$.

The negative value of $\dot{\mathcal{F}}_\text{BH}(E)$ for certain values of the energy $E$ is rather 
unusual.\footnote{One can also compute the transition rate in the 
AdS-Rindler patch of anti-de Sitter space. Independent on the boundary conditions, the transition rate is positive for any values 
of the energy $E$. However, the negative transition rate has been recently obtained in~\cite{Brenna&Mann&Martinez},
although in a completely different set-up.}
This may be due to either inapplicability of the
Gaussian approximation or impossibility of having a well-defined quantum field satisfying the Dirichlet boundary
condition on the boundary of the Schwarzschild--AdS geometry. It is worth noting that this is not the case in the
asymptotically flat space. Indeed, the thermodynamics of the Schwarzschild black hole in a spherical isothermal 
cavity is similar to a certain extent to that in asymptotically AdS space~\cite{York}. 
Assuming that the size of the cavity is much larger than
the Schwarzschild radius and imposing the Dirichlet boundary condition on the cavity, one finds that the transition
rate $\dot{\mathcal{F}}_\text{BH}(E)$ is non-negative near the cavity for any values of $E$. This is because 
$\Delta^{\frac{1}{2}}(\rho) \leq 1$ as it follows from \eqref{eq:square-root-delta} for vanishing cosmological constant, 
$\Lambda = 0$.

The response of the Unruh-DeWitt detector in Schwarzschild-AdS space has been numerically treated
in~\cite{NHLMMM}. The boundary condition imposed on the physical (non-rescaled) field $\Phi(x)$ is there of the Dirichlet 
type, while I have imposed this condition on the rescaled field $\bar{\Phi}(x) = \Phi(x)f^{\frac{1}{2}}(r)$. 
A linear combination of $\bar{\Phi}_\text{D}(x)/f^{\frac{1}{2}}(r)$ ($\bar{\Phi}_\text{D}(x)|_{\partial\mathcal{M}} = 0$) and 
$\bar{\Phi}_\text{N}(x)/f^{\frac{1}{2}}(r)$ ($\partial_r\bar{\Phi}_\text{N}(x)|_{\partial\mathcal{M}} = 0$) 
vanishes on the boundary $\partial\mathcal{M}$. That is their linear combination satisfies the Dirichlet condition as well.
Therefore, the transition rate computed in~\cite{NHLMMM} should not 
precisely match the transition rate found above.\footnote{However, $\dot{\mathcal{F}}_\text{BH}(E)$ oscillates with a 
frequency $\Omega$ as it can be seen 
in~\eqref{eq:transition-rate-total}. For $r_+ = 0.1$ and $r = 1$, the period of these oscillations $2\pi/\Omega$ approximately 
equals $4.82$. This is roughly the same as in~\cite{NHLMMM} after smoothing the plot of fig.~1 given in that paper.}

The part of the transition rate \eqref{eq:transition-rate-total} being due to the thermal gas in AdS space is non-negative for both 
types of the boundary conditions. The rate $\dot{\mathcal{F}}_\text{AdS}(E)$ is non-zero only for a discrete values of the 
energy $E$. This occurs due to the discreteness of the spectrum of the field excitations defined with respect to the Killing 
vector $K_1$. This is not the case, if one considers the Poincar\'{e} patch of anti-de Sitter space. Indeed, the Killing 
vector providing an automorphism of this patch into itself corresponds to the element $L_{05} + L_{03}$ of 
$\mathfrak{so}(2,3)$ which generates a continuous spectrum.

Thus, an observer located near the AdS boundary $\partial\mathcal{M}$ can discover quantum gravity by probing the
vacuum by the Unruh-DeWitt detector. The narrow peaks in $\dot{\mathcal{F}}(E)$ is an indicator of the AdS geometry,
while non-vanishing $\dot{\mathcal{F}}(E)$ between the peaks is due to the AdS black hole geometry.

\subsection{\bf Mean field squared $\Phi^2(x)$}

One might {\sl a priori} expect that no deviations in the value of $\Phi^2(x)$ should appear at a given space-time point in 
comparison with the Killing ansatz~\cite{Frolov&Zelnikov} for the large AdS black hole, i.e. $r_+ \gg 1$. For both 
the small black hole, $r_+ \ll 1$, and of the intermediate size, $r_+ \approx 1$, the Killing ansatz should not be appropriate, 
because both geometries are simultaneously relevant. At the zero temperature limit, the Schwarzschild--AdS geometry is thermodynamically and dynamically unstable, so that one might further await that $\hat{\Phi}^2(x)$ reduces to that
in anti-de Sitter space (this can be obtained by using findings of~\cite{Emelyanov}). However, these arguments should be
applicable if one does not impose the boundary conditions.

Indeed, substituting the total Wightman function \eqref{eq:wightman-total} in 
\eqref{eq:local-temperature-2squared-detector}, one obtains
\beqa\label{eq:phi-squared-total}
\Phi^2(x) &=& \frac{1}{24\pi^2} +
\frac{\gamma_\text{BH}}{2\pi^2f(r)}\left(\frac{\pi^2}{6\beta^2} + \frac{4f(r) - (f'(r))^2}{96}
\mp \frac{\Delta^{\frac{1}{2}}(2r_*)}{8r_*}\kappa\coth\kappa r_*\right)
\\[1mm]\nonumber&&
+ \frac{\gamma_\text{AdS}}{2\pi^2f_0(r)}
\left(\sum\limits_{n = 1}^{+\infty}\frac{n}{e^{\beta_0 n}-1} 
\mp \frac{1}{4}\sum\limits_{n =1}^{+\infty}\frac{1}{\sinh^2\left(\frac{\beta_0 n}{2}\right) + \sin^2r_*}
\mp \frac{1}{8\sin^2r_*}\right),
\eeqa
where, as above, the minus (plus) sign corresponds to the Dirichlet (Neumann) boundary condition.
Let us first discuss the formula \eqref{eq:phi-squared-total} omitting terms being due to the boundary conditions.
Then \eqref{eq:phi-squared-total} is applicable near the horizon of the black hole as well. The black 
hole term in \eqref{eq:phi-squared-total} does not diverge at the horizon if and only if $\beta = 1/T_\text{HP}$. 
This is analogous to the case of Schwarzschild--Minkowski space, wherein $\Phi^2(x)$ is also finite 
on the horizon only at the Hawking temperature~\cite{Frolov&Novikov}. At the AdS boundary, the second and third terms
in \eqref{eq:phi-squared-total} vanish and  $\Phi^2(x)$ approaches $1/24\pi^2$.

Restoring those (due to the boundary conditions) terms in \eqref{eq:phi-squared-total}, the mean field 
squared approximately becomes
\beqa\label{eq:mfs-on-boundary}
\Phi^2(x) &\approx& \frac{1}{24\pi^2} \mp \frac{1}{16\pi^2}
\eeqa
in the vicinity of the boundary $\partial\mathcal{M}$. Thus, an observer located near the AdS boundary 
cannot discover quantum gravity by using the mean field squared as a detector for probing the vacuum.
However, the mean field squared is sensitive to the type of the boundary conditions. This is not the case in
general for local observables. For instance, the stress tensor of the non-interacting scalar field conformally 
coupled to gravity in AdS space is insensitive to them~\cite{Emelyanov3}.

\section{Discussion}
\label{sec:Discussion and concluding remarks}

\subsection{Measuring spacetime topology in the semi-classical reime}

Among of the quantum gravity approaches (see~\cite{Kiefer} for a comprehensive review), it seems only the path 
integral method allows to have topologically inequivalent geometries at once. This approach was in particular used 
to discuss thermodynamical properties of the AdS black hole~\cite{Hawking&Page}. Depending on the size of
the black hole horizon $r_+$, the Schwarzschild--AdS geometry can be thermodynamically either stably or unstable.
For sufficiently small $r_+$, i.e. $r_+ < 1/3^{\frac{1}{2}}$, the AdS geometry filled with the thermal gas turns out
to be thermodynamically favored (as well as dynamically, see~\cite{Prestidge}). Despite of the thermodynamical
preference of one of these geometries, the quantum (for instance, scalar) field is simultaneously sensitive to the global 
structure of both geometries. This straightforwardly follows from the two-point function computed by employing the
path integral in the saddle-point approximation (see equation~\eqref{eq:feynman-function}). In this sense, 
quantum gravity reveals itself as simultaneously coexisting spacetimes with different topologies~\cite{Hawking2}. 

The global structure of topologically inequivalent geometries are in general rather different. Taking this into account, 
one can in principle construct a detector that could measure that quantum gravity effect even being in the semi-classical
regime. In the set-up considered in this paper, this is modeled by the Unruh-DeWitt detector. It turns out to be
possible, because the spectrum of the field excitations is discrete in AdS space ($\mathcal{M}_1$) and continuos in 
Schwarzschild--AdS space ($\mathcal{M}_2$). A detector modeled by the Wick squared field operator, $\hat{\Phi}^2(x)$, 
turns out to be insensitive to this effect. This can be accounted for its local probe of the vacuum. In the vicinity of 
$\partial\mathcal{M}$, i.e. $r \gg (2M)^{\frac{1}{3}}$, both geometries are almost identical and, thus, 
$\langle\hat{\Phi}^2(x)\rangle_{\mathcal{M}_1} \approx \langle\hat{\Phi}^2(x)\rangle_{\mathcal{M}_2}$. This should also 
be the case for the stress energy tensor computed close to $\partial\mathcal{M}$ in these spacetimes.

It should be emphasized that the results obtained above are valid if the Gaussian approximation is still a good
approximation for the two-point function in the vicinity of the AdS boundary.

It would be further interesting to find similar (due to different topologies) effects, but in less exotic set-ups.

\subsection{Algebras $\mathcal{A}(\mathcal{M}_2)$ and $\mathcal{O}(\mathcal{M}_2)$}

In the algebraic approach to quantum field theory (for instance, see~\cite{Haag,Hollands&Wald,Khavkine&Moretti}),
one works with the $*$-algebra of the local field operators. This algebra is ascribed to globally hyperbolic spacetime
endowed with a classical metric, i.e. this is a semi-classical approach.
In the set-up of this paper, I have implicitly worked with two such algebras, namely $\mathcal{A}(\mathcal{M}_1)$
and $\mathcal{A}(\mathcal{M}_2)$. 

According to the AdS/CFT correspondence, there should exist 
a map (injective $*$-homomorphism) of $\mathcal{A}(\mathcal{M}_1)$ into the algebra of the CFT operators
$\mathcal{O}(\partial\mathcal{M})$. In other words, the algebra $\mathcal{A}_1(\partial\mathcal{M})$ (obtained 
from $\mathcal{A}(\mathcal{M}_1)$ by pulling it to the boundary) should be isomorphic
to a certain subalgebra of $\mathcal{O}(\partial\mathcal{M})$.
The dynamics in $\mathcal{A}(\mathcal{M}_1)$ is set by an
automorphism $\alpha_{K_1}$ generated by the Killing vector $K_1 \in \mathfrak{so}(2,3)$. The state 
$|\Omega_1\rangle$, i.e. the AdS vacuum, is invariant under this algebra automorphism. Hence, 
$\alpha_{K_1}$ is unitary implementable on the algebra of the field operators.\footnote{More rigorously, on the GNS 
representation of $\mathcal{A}(\mathcal{M}_1)$ induced by that state (see~\cite{Haag} for details).} The time evolution 
is then governed by the self-adjoint operator $\hat{K}_1$ and, hence, is unitary. The same operator $\hat{K}_1$ plays 
a role of the Hamiltonian in $\mathcal{A}_1(\partial\mathcal{M})$ and, thus, the dual CFT theory evolves unitary as well.

One could instead start with the algebra $\mathcal{O}(\partial\mathcal{M})$. Then, 
applying the holographic principle, one associates with it a bulk algebra $\mathcal{O}(\mathcal{M}_1)$. 
The algebra $\mathcal{A}(\mathcal{M}_1)$ belongs then to $\mathcal{O}(\mathcal{M}_1)$ according to the
correspondence.
The mapping of $\mathcal{O}(\partial\mathcal{M})$ into $\mathcal{O}(\mathcal{M})$ can be 
understood as a bijective $*$-homomorphism between a certain theory in $\mathcal{M}_1$ and
a certain CFT theory in $\partial\mathcal{M}$~\cite{Rehren}. In practice, one achieves that through the so-called 
``smearing function'' which allows to obtain a bulk operator from a CFT operator defined on the 
boundary~\cite{Hamilton&Kabat&Lifschytz&Lowe}. A slightly different approach is advocated 
in~\cite{Papadodimas&Raju}
(see also~\cite{Banks&Douglas&Horowitz&Martinec,Balasubramanian&Kraus&Lawrence&Trivedi,Bena}).

This construction is expected to work for the AdS black hole geometry as well. One has 
that $\mathcal{A}(\mathcal{M}_2)$ is a subalgebra of the total algebra $\mathcal{A}(\mathcal{K})$ 
that can be ascribed to the Kruskal extension $\mathcal{K}$ of $\mathcal{M}_2$, such that
$\mathcal{A}(\mathcal{K}) = \mathcal{A}(\mathcal{M}_2){\otimes}\mathcal{A}(\mathcal{M}_2')$, where
$\mathcal{M}_2'$ is the causal complement of $\mathcal{M}_2$, i.e. an unavailable part of $\mathcal{K}$
for an observer inhabiting $\mathcal{M}_2$. A map $\alpha_{K_2}$ generated by $K_2$ is an automorphism 
of $\mathcal{A}(\mathcal{M}_2)$. The Hartle-Hawking state is invariant under $\alpha_{K_2}$. Thus, the time 
evolution is unitary on the Hilbert space representation of $\mathcal{A}(\mathcal{M}_2)$ defined with respect 
to this state. According to the AdS/CFT correspondence extended to $\mathcal{M}_2$, the algebra 
$\mathcal{A}(\mathcal{M}_2)$ should be isomorphic to a certain subalgebra of $\mathcal{O}(\partial\mathcal{M})$. 
The dynamics in $\mathcal{A}_2(\partial\mathcal{M})$ is still governed by the Hamiltonian $\hat{K}_2$. The crucial 
question is whether $K_2$ belongs to $\mathfrak{so}(2,3)$ when restricted to the boundary. 
If it does, then the boundary CFT theory isomorphic to $\mathcal{A}_2(\partial\mathcal{M})$ evolves unitary in
Schwarzschild time. Indeed, the dynamics set by $K_2$ is then unitary 
implementable on the CFT Hilbert space (built on $|\Omega\rangle$, see below).

There is no doubt $K_2$ is an element of $\mathfrak{so}(2,3)$. 
It is tempting to say that $K_2$ should be identified with the Killing vector 
$K_1$ as it follows from the form of the barred metric taken on the boundary.\footnote{Indeed, 
in the case $r \gg (2M)^{\frac{1}{3}}$, the optical metric approaches the line element of the three dimensional
Einstein static universe, i.e. $d\bar{s}^2 = dt^2 - d\theta^2 - \sin^2\theta d\phi^2$ at $r_* \rightarrow 0$. This is in
accordance with the boundary topology $\mathbf{R}{\times}\mathbf{S}^2$.}
However, this is in conflict with a certain expectation. Indeed, it is supposed that $\hat{K}_2$ is
the Hamiltonian for a set of the CFT operators isomorphic to $\mathcal{A}_2(\partial\mathcal{M})$. A two-point 
function in the (thermal) KMS state defined with respect to $K_2$ is then identical
to that for the the thermal gas in AdS space when restricted to $\partial\mathcal{M}$. Thus, the Hartle-Hawking 
state, i.e. $|\Omega_2\rangle$, is a ``thermalized'' version of $|\Omega_1\rangle$. This is incorrect.
Therefore, $K_2$ seems not to be equal to $K_1$.

This also can be justified by direct calculations. To show it, let us return to the scalar field $\Phi(x)$ conformally 
coupled to gravity in $\mathcal{M}_2$. On the boundary, the Wightman two-point function becomes
\beqa\label{eq:2pf-bh-boundary}
\langle\hat{\bar{\Phi}}(y)\hat{\bar{\Phi}}(y')\rangle &\propto& \frac{\kappa\sinh\left(\kappa\Theta\right)}{\sin\Theta}
\frac{1}{\cosh\left({\kappa\Theta}\right) - \cosh\left({\kappa\Delta t}\right)}
\eeqa
for the correlator of the form \eqref{eq:wightman-boulware-thermalized} (or the Neumann boundary condition), 
where $y,y' \in \partial\mathcal{M}$ and
$\cos\Theta = \cos\theta\cos\theta' + \sin\theta\sin\theta'\cos\left(\phi - \phi'\right)$.
In the no black hole case, i.e. $\kappa = i$, the correlator \eqref{eq:2pf-bh-boundary} reduces to the two-point 
function of the CFT operator with the conformal dimension  $\Delta_{-} = 1$ in the CFT 
vacuum~\cite{Balasubramanian&Kraus&Lawrence,
Banks&Douglas&Horowitz&Martinec,Balasubramanian&Kraus&Lawrence&Trivedi}. 
However, it is not evident for real values of $\kappa$ that there is a certain CFT operator of the conformal dimension 
$\Delta_{-} = 1$ which has \eqref{eq:2pf-bh-boundary} in a certain thermal state at temperature $\kappa/2\pi$.

The two-point function \eqref{eq:2pf-bh-boundary} corresponds to the KMS state defined with respect to 
$K_2$. It is expected that there is a pure (high-energy) CFT state, $|\Omega\rangle$, which should reproduce the 
two-point function \eqref{eq:2pf-bh-boundary} in the large $N$ limit~\cite{Banks&Douglas&Horowitz&Martinec}
(see also~\cite{Papadodimas&Raju}).
If one works in the semi-classical limit, then $|\Omega\rangle$ should be identified 
with the Hartle-Hawking state $|\Omega_2\rangle$.

Let us suppose that $|\Omega\rangle$ does exist with above mentioned properties. Then, one can mimic the black 
hole in AdS space at the quantum level. Indeed, one can find the positive- and negative-frequency modes with 
respect to $K_2 \in \mathfrak{so}(2,3)$ (assuming $K_2 \neq K_1$) in the AdS geometry. These scalar field modes 
determine a state which is equivalent to $|\Omega\rangle$. Thus, an observer in the vicinity of $\partial\mathcal{M}$ 
could claim by probing the scalar field vacuum that he is in the background of the black hole. Moreover, there should 
exist a coordinate patch of AdS space with the boundary topology $\mathbf{R}{\times}\mathbf{S}^2$ and 
the two-point function of the form \eqref{eq:2pf-bh-boundary} near $\partial\mathcal{M}$. This AdS patch 
has a horizon. For $|\Omega\rangle$ to be regular on the horizon, it must be the AdS vacuum state
restricted to this patch. Hence, $|\Omega\rangle$ has to be the ordinary CFT vacuum restricted to 
a certain patch of the AdS boundary.\footnote{This might be a clue to the resolution of the seemingly absent 
Poincar\'{e} recurrence.} In terms of the field operators, the CFT algebra isomorphic
to $\mathcal{A}_2(\partial\mathcal{M})$ should be related to $\mathcal{O}(\partial\mathcal{M})$
as $\mathcal{A}(M_2)$ related to $\mathcal{A}(\mathcal{K})$. The ordinary CFT vacuum probed by elements of 
this CFT subalgebra appears then as a thermal state at the Hawking-Page temperature. 

Thus, in general, a black hole formation could be envisaged in the semi-classical approximation as follows.
The observer has at his disposal
an algebra $\mathcal{A}_\text{obs}(\mathcal{M}) \subset \mathcal{A}(\mathcal{M})$, where $\mathcal{M}$
is a geometry representing the black hole formation in AdS space. He can use elements of 
$\mathcal{A}_\text{obs}(\mathcal{M})$ to probe the vacuum 
occupied by the scalar field. This vacuum is self-consistently ``chosen'' by the system, 
such that the semi-classical approximation is valid. The algebra $\mathcal{A}_\text{obs}(\mathcal{M})$ is 
supposed to change from $\mathcal{A}(\mathcal{M}_1)$ to $\mathcal{A}(\mathcal{M}_2)$.
The observer can ``define'' a vacuum and a particle associated with his dynamics in each of the asymptotic
cases (either $\mathcal{M}_1$ or $\mathcal{M}_2$). These concepts do not exist in general~\cite{Birrell&Davies}. 
A non-unitary evolution of observer's vacuum is then possible as 
Hilbert space representations of $\mathcal{A}_\text{obs}(\mathcal{M})$ can be unitary 
inequivalent (see, for instance,~\cite{Haag}). This does not, however, imply that the system's vacuum evolves
non-unitary. Thus, it seems the resolution of the information paradox~\cite{Hawking5} should 
be settled down if one appropriately prescribes the geometry during the final stage of the evaporation, when 
one cannot ignore the back-reaction of $\hat{\Phi}(x)$ on the geometry $\mathcal{M}_2$.

The above discussion is speculative, because it has not been proved that $|\Omega\rangle$ exists.
This non-trivial question is further investigated in~\cite{Emelyanov4}. Let us speculate further and assume 
$|\Omega\rangle$ does not actually exist with the needed properties. Then, the field theory considered in this 
paper has no holographic 
description.\footnote{In a recent paper~\cite{Papadodimas&Raju1}, the formula (4.20) is not proved.
The authors do not demonstrate that the scalar field modes $f_{\omega lm}(t,r_*,\Omega)$ exist, such those
the last formula in sec. 4.2.1 is indeed fulfilled for $|E\rangle$ ($\equiv |\Omega\rangle$).} 
This seems to be in agreement with conclusions of~\cite{Kabat&Lifschytz}. Besides, if $K_2$ does not coincide with
$K_1$, one cannot fit the boundaries of $\mathcal{M}_1$ and $\mathcal{M}_2$. That is $K_2$ must be $K_1$ 
(more on this in~\cite{Emelyanov4}).
In this sense, it would be interesting to investigate to which extent well-established results from the semi-classical 
quantum field theory in Schwarzschild--AdS black hole geometry characterized by $\mathcal{A}(\mathcal{M}_2)$ 
can be transferred to the theory $\mathcal{O}(\mathcal{M}_2)$.

\section*{
ACKNOWLEDGMENTS}

It is a pleasure to thank T. Erler, V. Frolov, I. Sachs, A. Vikman for discussions during preparation of this article.
I am especially thankful to M. Haack and D. Ponomarev for discussions 
and useful comments on an early version of this paper. I am also grateful to D. Sarkar for drawing my attention
to~\cite{Kabat&Lifschytz}. It is also a pleasure to thank an anonymous referee for comments/questions
which helped to improve the paper. This research is supported by TRR 33 ``The Dark Universe''.

\begin{appendix}
\section{Geodetic interval and the van Vleck-Morette determinant}
\label{app:Geodetic interval and the van Vleck-Morette determinant}

It is found convenient to compute these quantities using the Regge-Wheeler radial coordinate $r_*$. 
The optical (barred) metric then reads
\beqa
d\bar{s}^2 &=& dt^2 - dr_*^{2} - F(r_*)\big(d\theta^2 + \sin^2\theta d\phi^2\big)\,,
\eeqa
where $F(r_*) = r^2(r_*)/f(r_*)$ by definition. In the following, the argument of $F(r_*)$ is omitted
for the sake of transparency.

One computes them assuming points are close to each other, i.e.
$(r_* + \Delta r_*,\theta + \Delta\theta,\phi + \Delta\phi)$ and $(r_* ,\theta,\phi)$, where
$\Delta r_*, \Delta\theta$ and $\Delta\phi$ are of the same order of smallness. One has
\beqa\nonumber
\bar{\sigma}_3(\Delta r_*,\Delta\theta,\Delta\phi) &=&
\sum\limits_{n = 0}^{\infty}A_n(\Delta r_*)\big(\cos\Theta - 1\big)^n\,,
\eeqa
where
\beqa
\cos\Theta &=& \cos\theta\cos(\theta + \Delta\theta) + \sin\theta\sin(\theta + \Delta\theta)\cos(\Delta\phi)\,,
\eeqa
and
\beqa\nonumber
A_0(\Delta r_*) &=& \frac{1}{2}\Delta r_*^2 + \text{O}(\Delta r_*^7)\,,
\\[1mm]\nonumber
A_1(\Delta r_*) &=& 
-F - \frac{F'}{2}\Delta r_* + \frac{1}{12}\left(\frac{{F'}^2}{F}-2F''\right)\Delta r_*^2
- \frac{1}{24}\left(\frac{{F'}^3}{F^2} - 2\frac{{F'}F''}{F} + F^{(3)}\right)\Delta r_*^3
\\[1mm]\nonumber&&
+ \frac{1}{720}\left(19\frac{{F'}^4}{F^3} - 46\frac{{F'}^2F''}{F^2} + 16\frac{{F''}^2}{F}+
18\frac{F'F^{(3)}}{F} - 6F^{(4)}\right)\Delta r_*^4 + \text{O}(\Delta r_*^5)\,,
\\[1mm]
A_2(\Delta r_*) &=&
\frac{4F-{F'}^2}{24} - \frac{F'}{24}\big(F'' - 2\big)\Delta r_*
\\[1mm]\nonumber&&
-\frac{1}{720}\bigg(\frac{{F'}^4}{F^2} - 2\frac{{F'}^2}{F}\big(3F''-5\big)
-4\big(5-2F''\big)F'' + 9F'F^{(3)}\bigg)\Delta r_*^2 + \text{O}(\Delta r_*^3)\,,
\\[1mm]\nonumber
A_3(\Delta r_*) &=& -\frac{1}{720}\bigg(32F - \frac{{F'}^4}{F} - 10{F'}^2 + 3{F'}^2F''\bigg)
+ \text{O}(\Delta r_*)\,.
\eeqa
These allow to compute the three dimensional geodetic interval up to $\text{O}((x-x')^7)$.

Using an expression of $\bar{\sigma}_3(\Delta r_*,\Delta\theta,\Delta\phi)$, one can compute the square 
root of the van Vleck-Morette determinant up to the order $(x-x')^5$. The result is
\beqa
\Delta^{\frac{1}{2}}(\Delta r_*,\Delta\theta,\Delta\phi) &=& \sum\limits_{n = 0}^{\infty}B_n(\Delta r_*)\big(\cos\Theta - 1\big)^n,
\eeqa
where
\beqa\nonumber
B_0(\Delta r_*) &=& 1 + \frac{1}{24}\bigg(\frac{{F'}^2}{F^2}- 2 \frac{F''}{F}\bigg)\Delta r_*^2
-\frac{1}{24}\bigg(\frac{{F'}^3}{F^3}-2\frac{F'F''}{F^2}+\frac{F^{(3)}}{F}\bigg)\Delta r_*^3
\\[1mm]\nonumber&&
+\frac{1}{5760}\bigg(223\frac{{F'}^4}{F^4} - 532\frac{{F'}^2F''}{F^3} + 172\frac{{F''}^2}{F^2}
+216\frac{F'F^{(3)}}{F^2}-72\frac{F^{(4)}}{F}\bigg)\Delta r_*^4 + \text{O}(\Delta r_*^5)
\,,
\\[1mm]
B_1(\Delta r_*) &=&
\frac{F''-2}{12}+\frac{F^{(3)}}{24}\Delta r_*
-\frac{1}{1440}\bigg(4\frac{{F'}^4}{F^3}-5\frac{{F'}^2}{F^2}\big(3F''-2\big)
\\[1mm]\nonumber&&
+2\frac{F''}{F}\big(7F''-10\big)
+12\frac{F'F^{(3)}}{F} - 18F^{(4)}\bigg)\Delta r_*^2 + \text{O}(\Delta r_*^3)\,,
\\[1mm]\nonumber
B_2(\Delta r_*) &=&\frac{1}{1440}\bigg(76+5\frac{{F'}^4}{F^2}-14\frac{{F'}^2F''}{F}
+\big(9F''-40\big)F'' + 12F'F^{(3)}\bigg)+ \text{O}(\Delta r_*)\,.
\eeqa

\end{appendix}

\end{document}